# Nano-engineered Diamond Waveguide as a Robust Bright Platform for Nanomagnetometry Using Shallow Nitrogen Vacancy Centers

S. Ali Momenzadeh[†,*], Rainer J. Stöhr[†,**], Felipe Favaro de Oliveira[†], Andreas Brunner[†], Andrej Denisenko[†], Sen Yang[†], Friedemann Reinhard[†], and Jörg Wrachtrup[†,‡]

† 3rd Institute of Physics, Research Center SCoPE and IQST, University of Stuttgart, 70569 Stuttgart, Germany

‡ Max Planck Institute for Solid State Research, Heisenbergstrasse 1, 70569 Stuttgart, Germany

**ABSTRACT:** Photonic structures in diamond are key to most of its application in quantum technology. Here, we demonstrate tapered nano-waveguides structured directly onto the diamond substrate hosting shallow-implanted nitrogen vacancy (NV) centers. By optimization based on simulations and precise experimental control of the geometry of these pillar-shaped nano-waveguides, we achieve a net photon flux up to ~ $1.7 \times 10^6$ /s. This presents the brightest monolithic bulk diamond structure based on single NV centers so far. We observe no impact on excited state lifetime and electronic spin dephasing time ($T_2$) due to the nanofabrication process. Possessing such high brightness with low background in addition to preserved spin quality, this geometry can improve the current nanomagnetometry sensitivity ~ 5 times. In addition, it facilitates a wide range of diamond defects-based magnetometry applications. As a demonstration, we measure the temperature dependency of $T_1$ relaxation time of a single shallow NV center electronic spin. We observe the two-phonon Raman process to be negligible in comparison to the dominant two-phonon Orbach process.

Diamond defect centers are exquisite nanoscale sensors for a variety of physical parameters like magnetic[1] and electric[2] fields and temperature[3]. Among other parameters their sensitivity relies on proximity and photon detection efficiency. Nuclear magnetic resonance (NMR) experiments were recently shown[4-8] using the negatively-charged nitrogen vacancy (NV) center positioned few nanometers below the diamond surface ("shallow" NV center). These experiments became possible by the ability to optically address and readout spins of the NV center[9]. Yet, a major drawback of all magnetometry-based experiments with shallow NV centers is the low number of collected photons which causes long measurement times. At the same time, the photon count rate (F) also limits the magnetometry sensitivity[10] as it scales with $1/\sqrt{F}$.

A major reason for the low signal strength is the high refractive index mismatch[11] between diamond ($n_{diamond}$ ~ 2.4) and the collection medium (e.g. air; $n_{air}$ ~ 1). This causes most of the emitted photons from the NV center to be reflected back at the diamond-air interface into the diamond substrate. Even by benefiting from high NA microscope objective lenses, collection efficiencies are typically[12] below 10% resulting in total instrument detection efficiency on the order of 1%. Recently, efforts have been made to overcome this drawback by fabricating photonic structures directly onto the diamond surface in order to enhance the collection efficiency of either deep[12-17] or shallow NV centers[18]. Maletinsky *et al.*[18] demonstrated a monolithic diamond scanning tip based on nanopillars hosting NV centers ~ 10 nm below the nanopillars top surface. Such tips were fabricated from diamond nanopillars with a uniform diameter of ~ 200 nm and a height of >1 µm on a ~ 3 µm-thick diamond membrane. In this approach, the emission of the shallow NV center is coupled to the fundamental $HE_{11}$ mode of the waveguide which is guided back and collected by the microscope objective lens located below the membrane. To optimize for the best sensitivity of this nanostructure, we chose to increase the photon count rate while maintaining the spin coherence properties unaffected by the nanofabrication process. Here, we demonstrate a fluorescence rate up to ~ $1.7 \times 10^6$ photons/s from a single NV center resulting from 2.5 keV energy of nitrogen implantation located below the top facet of conically tapered nanopillar with top diameter ~ 400 nm, bottom diameter ~ 900 nm, and height of ~ 1.2 µm. Most importantly, due to the relatively large top diameter of the nanopillar (compared to previously shown diameter[18] of ~ 200 nm) destructive influences caused by fabrication, e.g. on the $T_2$ spin dephasing time, are minimized. These two im-portant features can improve the proton detection sensitivity ~ 5 times in comparison to the current sensitivity based on NV centers under the non-structured diamond surface[8]. All our waveguides were fabricated onto diamond substrates of ~ 30 µm thickness for which no special handling compared to bulk samples is required.

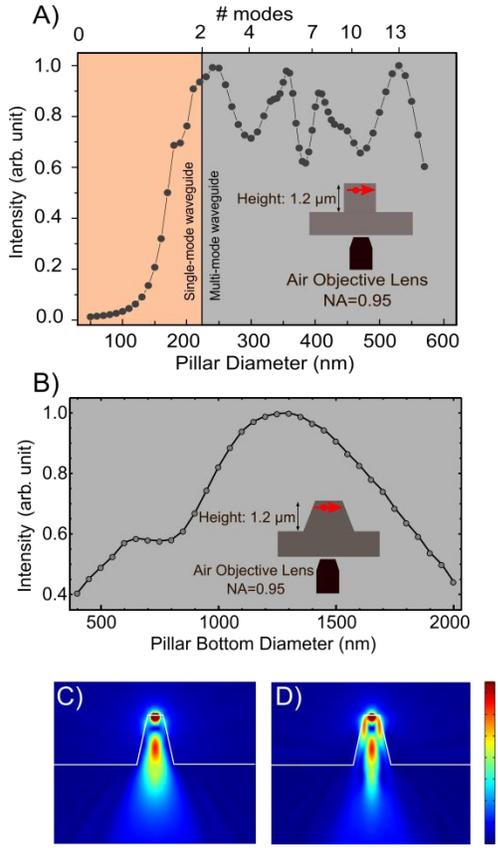

**Figure 1.** Optimization of the collected photon intensity by numerical simulations; the simulation of the electric field pattern of a dipole oriented perpendicular to the diamond waveguide axis radially located at its center with the depth of 5 nm below the surface, emitting light at λ=637 nm wavelength using 3D finite element method. (A) Light intensity emitted from the dipole collected by the objective lens as a function of the waveguide diameter for the case of a cylindrical waveguide. (B) For the case of a conical waveguide, the top diameter is fixed to 400 nm and the bottom diameter is varied to optimize towards the highest collected intensity achievable by the nanopillar waveguide geometry. (C) The in-plane and (D) out-of-plane emission intensity pattern of the above-mentioned electric dipole located inside a 400 nm – 900 nm waveguide with a height of ~ 1.2 µm.

Using 3D finite element simulations, we numerically calculated the influence of the nanopillar geometry on the number of emitted photons. The emission from a 5 nm-deep NV center is collected by an NA=0.95 microscope objective lens located below the diamond substrate on the opposite side of the nanopillar. The height of the waveguide was set to be 1.2 µm which is the upper reliable limit for our fabrication (see supporting information for more details). Figure 1A shows the collected intensity for a purely cylindrical waveguide of 1.2 µm height as a function of its diameter. For this case, the number of modes supported by the waveguide is given by:

$$\# \text{ modes} \approx (4/\pi^2) V^2 \qquad (1)$$

with V given by:

$$V = (2\pi/\lambda) (n^2_{diamond} - n^2_{air})^{0.5} \, r, \qquad (2)$$

where λ is the emission wavelength, $n_{diamond}$ and $n_{air}$ are the refractive indexes of diamond and air, respectively, and r is the radius of the waveguide[19]. From equations 1 and 2 as well as from the numerical results shown in Figure 1A it becomes obvious that for diameters below ~ 150 nm, the waveguide does not support any mode and hence no light is guided towards the objective lens. In this case, the lack of a mode for small diameters is because of the scattering induced by the diamond/air interface. Yet, for diameters close to ~ 200 nm the collected intensity reaches its maximum coinciding with a transition of the waveguide from single-mode to multi-mode. For larger diameters, the collected intensity exhibits a slight modulation corresponding to the resonances of the modes with respect to the diameter of the waveguide[20]. We optimized the performance of the nanopillar geometry by varying its bottom diameter independently from the top diameter. This variation modifies the effective refractive index and as a result, the propagation constant for each mode[21]. Having the propagation constants optimized, the highest collection efficiency of the emitted photons through the structure towards the objective lens can be achieved. As shown in Figure 1B, with the top diameter fixed at 400 nm, the maximum number of collected photons can be achieved from a nanopillar with bottom diameter of 1.2-1.3 µm. It should be mentioned that the emission spectrum of the NV center is relatively broad. The effect of emission wavelength on the simulated collection efficiency for an NV center located 5 nm below the top surface of a nanopillar with top diameter of 400 nm, bottom diameter of 900 nm, and height of 1.2 μm is presented in supporting information.

To realize this geometry experimentally, nitrogen ions ($^{15}N^+$) were implanted into a (100)-oriented ultrapure electronic grade ($^{13}C$ natural abundance) CVD diamond substrate of ~ 30 μm thickness with a dose of 100-200 ions/μm² at energies of 2.5, 5, and 10 keV. Based on the stopping and range of ions in matter (SRIM[22]) simulations, the NV centers can be approximately in the range of 3-20 nm of depth. The top-down fabrication method combining dedicated electron beam lithography (EBL) with reactive ion etching – inductively coupled plasma (RIE-ICP) recipes was utilized to transfer the tapered waveguide geometry onto the diamond substrate. A negative electron beam resist (FOX®25 from Dow Corning) was chosen as a mask against $O_2/O_2+CF_4$ plasma. Top and bottom diameters of the nanopillars were independently controlled by tuning the EBL exposure dose of the resist and simultaneously the corrosion rate of the mask by adjusting the plasma parameters. In this way, bottom diameters up to approximately 900 nm with fixed top diameter and height were achieved. Larger diameters did not show good resistance against etching (see supporting information for more details).

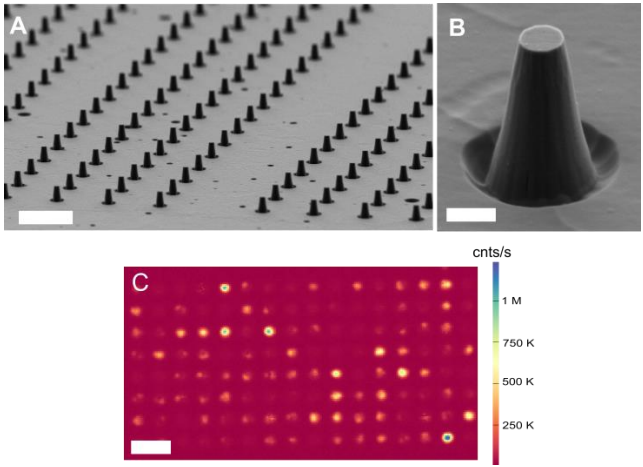

**Figure 2.** (A) SEM image of an array of tapered nanopillars (scale bar = 3 μm). (B) SEM image of a diamond tapered nanopillar; height ~ 1.2 μm, top diameter ~ 400 nm, bottom diameter ~ 900 nm (scale bar = 500 nm). (C) Confocal image of the tapered nanopillars (2.5 keV energy of implantation) under ambient conditions (scale bar = 5 μm).

Scanning electron microscopy (SEM) images of these structures are shown in Figures 2A-B. The optical properties of these waveguides were investigated by a home-built confocal microscope running under ambient conditions using excitation light of 532 nm wavelength and an air objective lens with NA=0.95 located on the non-structured side of the sample. Second-order autocorrelation measurements ($g^{(2)}$) were performed by detecting the fluorescence with two avalanche photodiodes in Hanbury Brown and Twiss (HBT) configuration. Figure 2C shows the confocal image of tapered-nanopillars with 400 nm of top and 900 nm of bottom diameter. This image illustrates the relatively low background of our geometry in contrast to previously demonstrated nanopillars with ~ 200 nm uniform diameter[16]. For instance, even at ~ 700 μW of incident laser power, an empty nanopillar has only ~ 15% of the count rate of a nanopillar with single NV center. In this sample, about 30% of the nanopillars showed a second-order autocorrelation signal indicating single photon emission with excellent photostability (see supporting information for more details).

In case of a single photon source inside a photonic structure the collected signal depends[23] on the collection efficiency of the structure and its spontaneous emission rate, which inversely depends on the excited state lifetime. To study the efficiency of our photonic structures, excited state lifetime and saturation behavior of the shallow NV centers were investigated. The excited state lifetime was measured using a 10ps-pulsed laser of 532 nm wavelength for shallow NV centers under the nanopillar facet as well as the non-structured diamond surface. Common to all, excited state lifetime of shallow NV centers were observed to be longer in comparison to the ones with depth of few microns[24]. This can be caused by a reduction of the local density of states (LDOS) due to the close vicinity to the diamond-air interface[25]. Nevertheless, no difference was observed between the excited state lifetime of shallow NV centers under the nanopillar surface and under the non-structured surface (see supporting information). Therefore, we conclude that our nanopillar fabrication method has no significant influence on the emission rate of the NV centers. To investigate the saturation behavior of the tapered nanopillars, we measured the number of detected photons as a function of the incident laser power. An enhancement (reduction) in the saturation count rate (power) of the NV center due to the nano-waveguide by more than one order of magnitude was achieved. Another noticeable point of this structure is the low background signal. The purity degree of our signal can be expressed via normalized photon shot noise $\sqrt{F_{total}}/F_{net}$ where $F_{total}$ and $F_{net}$ are the total and net detected fluorescence count rates, respectively. For instance, at P=1 mW this is ~ $9\times10^{-4}$ $\sqrt{Hz}$ for one of the investigated nanopillar NV centers. We also measured the saturation power and count rate of the NV centers resulting from 2.5 keV energy of ion implantation in this cone-shaped nanopillar geometry with top diameter fixed at ~ 400 nm and different bottom diameters. As shown in Figure 3A, the geometry with larger bottom diameter (~ 900 nm) is a more efficient waveguide for the excitation and emission of the NV centers in comparison to the ones with smaller bottom diameters. Presented in Figure 3A, this geometry exhibits saturation count rate up to ~ $1.7 \times 10^6$ photons/s. Showing no influence on the excited state lifetime while accompanied with a huge enhancement in the number of collected photons, we conclude that the collection efficiency of the NV center due to our nanostructure is highly augmented. Figure 3B highlights the reasonable agreement between the experimental values and the simulated intensity.

[3]

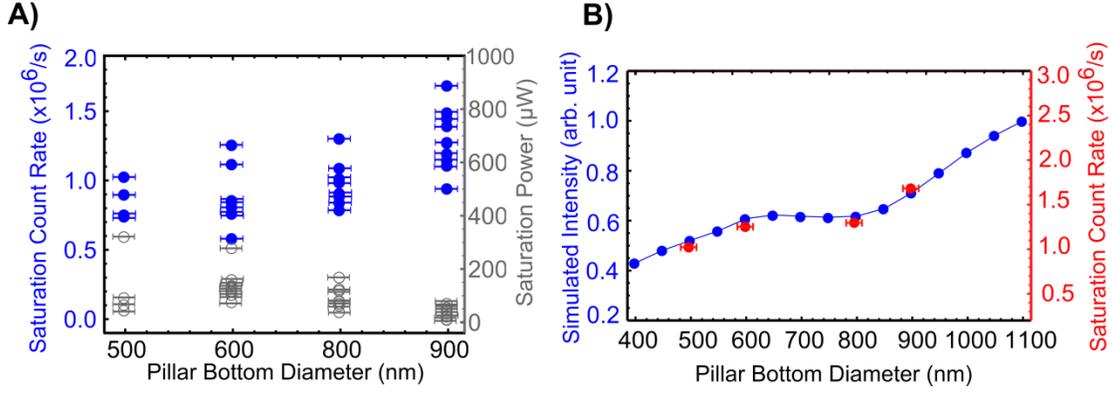

**Figure 3.** (A) Statistical study of saturation power (hollow gray circles) and saturation count rate (bold blue circles) of NV centers inside the nanopillars with respect to various bottom diameters when top diameter and height are fixed at 400 nm and 1.2 µm, respectively. Saturation count rate up to ~ $1.7 \times 10^6$ /s is measured from a nanopillar with bottom diameter ~ 900 nm. (B) Comparison of simulated and experimental results for different nanopillar bottom diameters while top diameter and height are kept to the above-mentioned values. The error bar is ±20 nm given by precision of our fabrication process and SEM imaging accuracy.

Besides improved collection efficiency, it is shown that the large diameter of the nanopillars also yields to preservation of the spin dephasing times of shallow NV centers. $T_2$ dephasing times were measured by means of optically detected magnetic Hahn-echoes probing the coherence of the ground state $m_s=|-1\rangle$ to $m_s=|0\rangle$ of the NV center electronic spin under ambient conditions in the presence of an external magnetic field of 100-150 G parallel to the NV axis. The $T_2$ times from the NV centers resulting from 2.5 keV energy of ion implantation within our nanopillar structure were obtained up to ~ 23 µs similar to the ones under the non-structured diamond surface. We also performed dynamical decoupling scheme to enhance the $T_2$ time of the NV centers inside the nanopillar. As an example, using CPMG pulse scheme with 80 π-pulses, we could extend the $T_2$ time up to ~ 90 µs for the NV centers resulting from 2.5 keV of implantation energy. Based on our statistical study, we conclude that our geometry also preserves the spin properties of the shallow NV centers (further information about the values of $T_2$ times for different implantation energies are provided in supporting information). Improved surface preparation techniques will help to further improve the $T_2$ dephasing times[26,27].

Recently, low temperature (LT) magnetometry has emerged as a new branch in the field of NV-based metrology[27-30]. At LT, the $T_2$ coherence time of NV ensemble was shown to be enhanced using dynamical decoupling schemes approaching half of the spin-lattice relaxation time[30] $T_1$. $T_1$ relaxation times of an ensemble of NV centers at LT were reported[30,31] exceeding one minute. With the aid of high collection efficiency of the presented nanopillar structure, we are reporting the value of the $T_1$ relaxation time of a single shallow NV center at room temperature down to ~ 5K. Using the same pulse scheme from Ref. 31, we measured $T_1$ values of 95.3±9.94 ms and 306.8±49.8 ms, in the absence and presence of applied magnetic field (~ 100 G || NV axis), respective-

ly, at ~ 5K. In comparison to the former reported values from ensembles of NV centers[31], this NV center shows a shorter $T_1$ time. This can be caused by magnetic noises at its surrounding, in addition to other surface defects. By applying magnetic field of ~ 100 G aligned to the axis of the NV center, we could extend the $T_1$ time by more than three times (Figure 4, inset). This might be due to decoupling from S=1/2 spin defects like P1 centers[31]. We measured the $T_1$ relaxation time of this single NV center versus different temperatures in the absence of applied magnetic field. As shown in Figure 4, as the temperature is decreased from room temperature down to ~ 5K, the $T_1$ relaxation time increases from ~ 5 ms to ~ 100 ms.

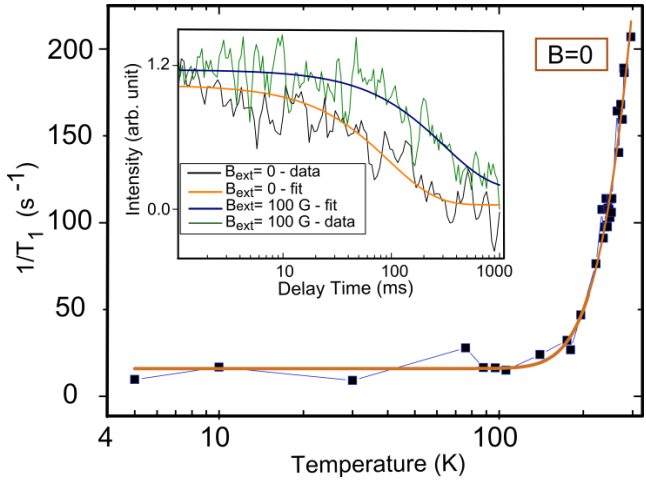

**Figure 4.** Inset: $T_1$ relaxation time of a single shallow NV center inside the tapered nanopillar geometry at ~ 5K in the absence and presence of an external 100 G aligned magnetic field. Main: $T_1$ relaxation time vs. temperature in the absence of applied magnetic field for the same shallow NV center.

This dependency can be considered as a sum of a sample-dependent term and a two-phonon Orbach process term:

$$\frac{1}{T_1} = A + \frac{B}{\exp\left(\frac{\Delta}{K_B T}\right) - 1} \quad (3)$$

where A and B are fitting parameters, $\Delta$ is the dominant vibrational energy, and $K_B$ is the Boltzmann's constant[31]. In comparison to Ref. 31, the Orbach process was observed to be the only temperature-dependent term, while the two-phonon Raman process could be negligible[31,32]. Fitting the data shown in Figure 4 to equation 3, the vibration energy of $\Delta$=92.92±6.40 meV is obtained. The difference of this value in comparison to previously shown data[31], ~ 70 meV, can be related to the modified phonon density of states[33] for this shallow NV center, or the presence of paramagnetic noises at its surrounding. Both can be attributed to the close vicinity of the NV center to the diamond surface.

In conclusion, we showed a reliable nanostructure for hosting shallow-implanted NV centers using subtle nanofabrication techniques. A photon count rate ~ $1.7 \times 10^6$ /s was achieved which sets a new record value for NV centers in monolithic bulk diamond structures. Due to its geometry, no degradation of the $T_2$ dephasing time was observed in comparison to NV centers below the non-structured diamond surface. We concluded that the spin properties of the NV centers are preserved throughout the whole nanofabrication process. Finally, we presented application of the nanostructures as a powerful platform for LT magnetometry-based measurements. We reported the $T_1$ relaxation time of a single NV center at LT exceeding 300 ms. Using this structure, further investigations on the phononic environment of the shallow NV center were also reported. This tapered nanopillar geometry can serve as a powerful enabling tool towards achievement of more detailed information regarding the shallow NV centers, besides further LT as well as ambient condition NMR-based measurements using the NV center probe.

## AUTHOR INFORMATION

**Corresponding Author**
* E-mail: a.momenzadeh@physik.uni-stuttgart.de
** E-mail: r.stoehr@physik.uni-stuttgart.de

**Notes**
The authors declare no competing financial interest.

## ACKNOWLEDGMENT

We acknowledge I. Gerhardt, P. Neumann, T. Staudacher, M. Doherty, R. Kolesov, and M. Nesterov for fruitful discussions. The great hospitality and support of Nano Structuring Lab headed by Jürgen Weiss at Max Planck Institute for Solid State Research is very much appreciated. S.A.M is especially grateful to K. Panos and K. Lindfors for EBL training. F.F. appreciates the financial support by CNPq project number 204246/2013-0. A. B. acknowledges the financial support by Hans L. Merkle-Stiftung. The authors would like to acknowledge financial support by the EU via projects SQUTEC, DIADEMS, SIQS, and QINVC; the DFG via Research Group 1493; the BMBF via Q.Com; and the Max Planck Society.

## ABBREVIATIONS

NV, nitrogen vacancy; EBL, electron beam lithography; RIE-ICP, reactive ion etching - inductively coupled plasma; LT, low temperature

# Supporting Information

1) **Fabrication process**

The substrates are single crystalline ultrapure (100)-oriented diamond ($^{13}$C natural abundance) with thickness of ~ 30 μm provided by Element Six and Delware Diamond Knives (DDK). After nitrogen implantation, annealing, and acid boiling, the substrate was glued by PMMA on top of a silicon wafer for structural support. 5 nm of Cr was thermally evaporated on top of the substrate as an adhesion layer. After spin coating and soft baking of FOX®25 (the resulting thickness ~ 450 nm), another 5 nm layer of Cr was deposited as an anti-charging layer. EBL was done in Raith Eline apparatus with 20 kV of acceleration voltage and 10 μm aperture with an exposure dose of 3000-4000 μC/cm$^2$. After exposure, anti-charging Cr layer was chemically etched. The exposed mask was developed by MF-322 developer. RIE-ICP process was performed in Oxford PlasmaPro NGP80 machine at initial sample condition at room temperature under vacuum of P < 1.0×10$^{-6}$ mbar. Oxygen plasma (30 sccm) was used in three steps with duration of each 2 minutes using constant RIE power of 100 W under 10 mTorr chamber pressure. The ICP power was adjusted up to 600W to control the corrosion rate of the mask and thus the nanopillar shape. Between each consequent oxygen etching step, 7 seconds of O$_2$/CF$_4$ (30 sccm O$_2$ / 2 sccm CF$_4$) with the RIE and ICP powers of 30W and 150W, respectively, was used to etch out the residual FOX®25 particles located on the bare surface of diamond [SI-1]. This short step was observed to be enough to avoid formation of any diamond nanograss. The whole etching process results in an etching selectivity of ~ 6. After the final etching step, the residual of the mask (thickness ~ 200 nm) was removed by 20 minutes immersion in buffered hydrofluoric acid (BHF) solution. Finally, samples were once more boiled in triacid mixture (H$_2$SO$_4$:HNO$_3$:HCLO$_4$; 1:1:1 volume ratio) and cleaned by demineralized water, pure isopropanol, and dried by nitrogen.

2) **Optical Characterization**

The optical characterization of the nanostructures was done by a home-built confocal microscope running under ambient conditions using a 532 nm wavelength excitation laser. The collected phonon sideband is spectrally separated from the excitation light by a 650 nm long pass fluorescence filter after being spatially filtered by a 50 μm pinhole. To decide on the right size of the pinhole in the confocal setup, the beam collected back from the nanopillars by the microscope objective lens (Olympus MplanApo N 50x/0.95) was focused onto the chip of a cooled-CCD camera (Cascade 512B). The FWHM of the Airy curve was observed to be about 2 pixels large (each pixel: 16 μm). Therefore, a 50 μm pinhole was chosen to collect all signal photons while rejecting the major part of the background contribution. When considering the NV center as a three level system under relatively low incident power, neglecting dark counts of the detectors, the power-dependent detected fluorescence scales with power as[SI-2]:

$$F(p) = F_{sat}/(1+\frac{P_{sat}}{P}) + \text{Const.}P, \tag{SI-1}$$

where $F_{sat}$ and $P_{sat}$ are the saturation count rate and power, respectively. P is the incident power and Const is the fitting parameter. The last term on the right side of equation SI-1 is the linear power-dependent background signal. As an example, the net signal vs. pumping power of one of the single NV centers resulting from 2.5keV of ion implantation energy inside the tapered nanopillar (top diameter: ~ 400 nm, bottom diameter: ~ 900 nm, height: ~ 1.2 µm) is shown in Figure SI-1A together with its background signal and data collected for an NV center under the non-structured diamond surface for comparison. Based on equation SI-1, the saturation count rate and saturation power of the nanopillar NV center are found to be 1.36±0.01 Mcnts/s and 27±1 µW, respectively. For the NV center under the non-structured surface the values are found to be 132±0.7 kcnts/sec and 923±11 µW, respectively. The inset of Figure SI-1A depicts a normalized fluorescence second-order autocorrelation ($g^{(2)}$) signal from this nanopillar NV center clearly showing a single-photon source emission.

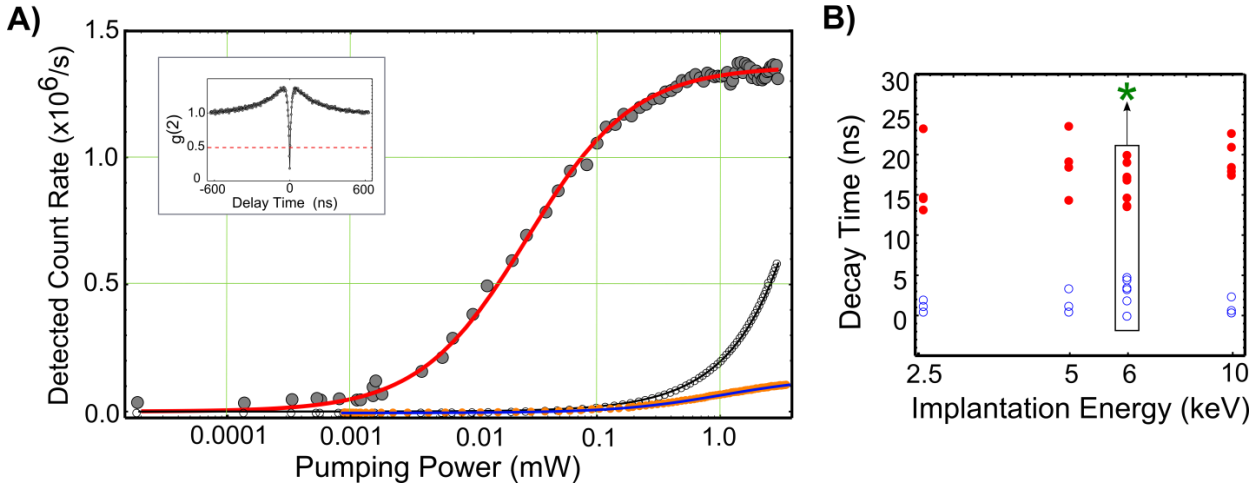

**Figure SI-1.** (A) Saturation curve for a single NV center inside the tapered nanopillar structure (bold gray circles with red fit) in comparison to the NV center under the non-structured diamond surface (bold orange circles with blue fit) and the deducted background signal of the nanopillar structure (black hollow circles with black fit). The inset shows a normalized fluorescence second-order autocorrelation signal showing a single-photon source emission from the NV center under the nanopillar surface. (B) Double-exponential decay times from the excited state lifetime signals for shallow NV centers. Hollow blue circles are shorter than 5 ns which can be attributed to unbleachable background contributions or semi-bleached NV centers in the vicinity of the probed NV center. Bold red circles are the longer decay rates, resampling the NV centers excited state lifetime, and are observed to be longer than 12 ns. For comparison, several NV centers under the non-structured diamond surface were inspected, which are shown in a rectangle marked with green star (*).

The excited state lifetime measurements were performed using 10 ps pulses of 532 nm wavelength laser together with time-correlated single photon counting electronics. For comparison, measurements were taken for NV centers under nanopillar surface as well as under the non-structured diamond surface. In all cases, a double exponential decay behavior was observed. By integration, the faster decay component shows less than 5% of signal contribution in comparison to the slower decay component. As can be seen in Figure SI-1B, the fast decay rate was observed to be shorter than 5 ns (hollow blue circles) consistently. We attribute this contribution to an unbleachable background or partially quenched NV centers in the vicinity of the probed NV center. However, the longer decay rate, resampling the excited state lifetime, was observed to always exceed 12 ns which is the value reported for NV centers located few microns below the diamond surface[SI-3]. The increase on the excited state lifetime can be because of a reduction of the local density of states (LDOS) due to the close vicinity to the diamond-air interface[SI-4].

### 3) Coherence time measurements

$T_1$ and $T_2$ measurements were accomplished by means of optical and microwave pulses generated by programmed FPGAs. The $T_2$ time was measured by means of optically detected magnetic resonance technique based on the Hahn-echo scheme probing the coherence of the ground state $m_s=|-1\rangle$ to $m_s=|0\rangle$ of the NV center electronic spin. Figure SI-2 shows the values obtained for NV centers resulting from different implantation energies all located within the same geometry of tapered nanopillars, and under the non-structured diamond surface. All measurements were done under ambient conditions in the presence of an external magnetic field (100-150 G) parallel to the NV axis. The values of the $T_2$ time were obtained from the Hahn-echo signal by the fitting equation mentioned in Ref. SI-5. As shown in Figure SI-2, the $T_2$ times associated with the NV centers under the non-structured diamond surface and nanopillar surface are similar. This means that the nanostructure preserves the spin properties of shallow NV centers even for 2.5 keV of ion implantation energy.

The measurement of $T_1$ relaxation time was performed in attoLIQUID1000 system (AttoCube) with a base temperature of ~ 5K. To avoid any undesired leakage of microwave and optical pulses, two cascaded accusto-optic modulators and three high-isolation microwave switches were used.

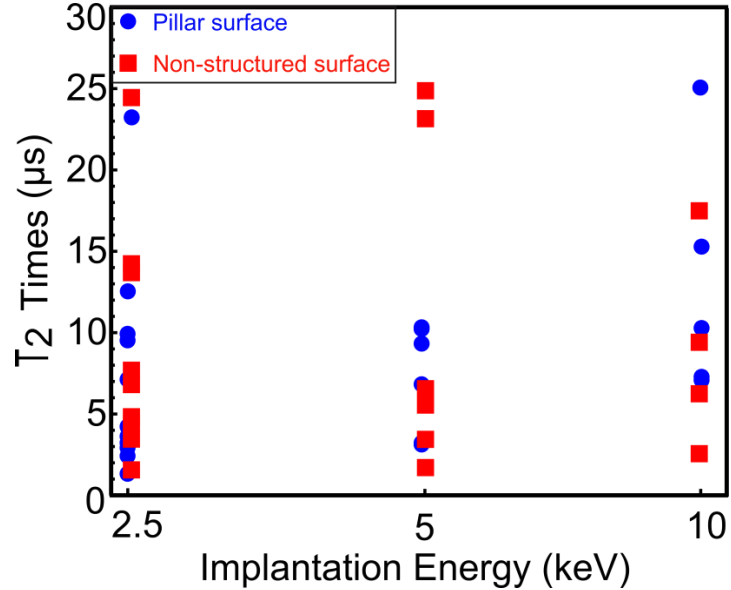

**Figure SI-2.** Hahn-echo $T_2$ times for NV centers under the diamond nanopillar (blue circles) and under the non-structured diamond surface (red squares) are shown vs. nitrogen implantation energy. No systematic difference between the values of NV centers' $T_2$ times in the two investigated geometries is observed. This indicates the minimized damage to the NV center due to the nanofabrication process.

4) **Wavelength-dependent collection efficiency**

The NV center in diamond has a broad emission spectrum, namely zero-phonon line (ZPL) at $\lambda \sim$ 637nm and phonon sideband (PSB) from ~ 650nm to ~ 800nm. In order to limit the computational effort, we performed all calculations for a single wavelength (i.e. the ZPL) only.

However, we have verified the collection efficiency of the 400nm/900nm pillar geometry as a function of wavelength ($\eta(\lambda)$) in the spectral range of the NV center emission. As shown in Figure SI-3, the curve has been normalized in a way such that the relative collection efficiency at the ZPL is set to unity. A decrease of approximately 15% in collection efficiency from $\lambda$=650nm to $\lambda$=800nm wavelength can be observed.

However, one has to keep in mind that also the NV center emission drops significantly for wavelengths longer than 700nm. By weighting the collection efficiency curve shown in Figure SI-3 with the NV center emission spectrum (also normalized to the ZPL emission) we determined that by performing the calculations for $\lambda$=637nm only, we overestimate the overall collection efficiency by only about 5%.

We therefore argue that simplifying the calculations to the single wavelength case of λ=637nm is a valid simplification of the real-world scenario.

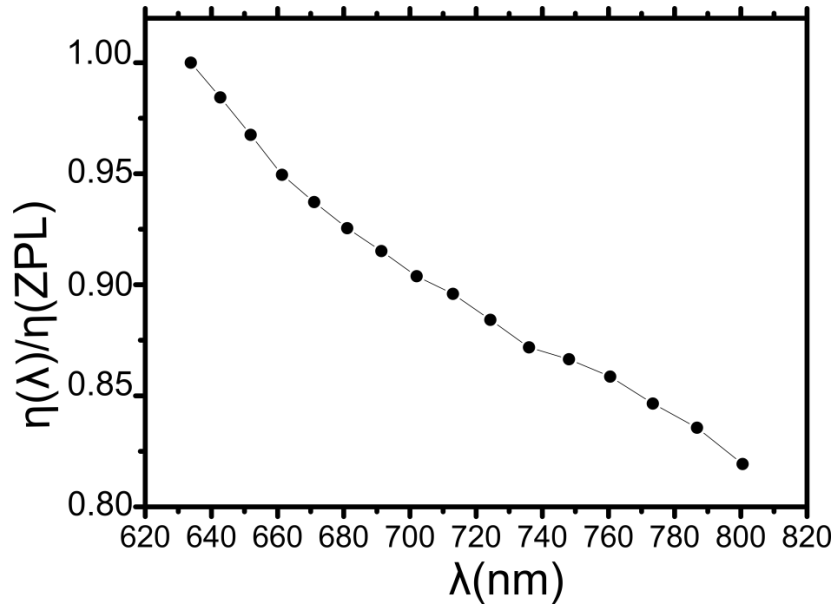

**Figure SI-3.** Normalized collection efficiency of the cone-shaped nanopillar geometry (top diameter: ~ 400nm, bottom diameter: ~ 900 nm, height: ~ 1.2 μm) simulated for an NV center located 5 nm below its top facet for different emission wavelength. As shown, the collection efficiency shows about 15% of decrease while the emission wavelength increases from 650nm to 800nm.